\documentclass[12pt]{article}
\usepackage[dvips]{graphicx}

\newcommand{\be}{\begin{equation}} \newcommand{\ee}{\end{equation}}
\newcommand{\ba}{\begin{eqnarray}} \newcommand{\ea}{\end{eqnarray}}
\newcommand{\nn}{\nonumber} \renewcommand{\bf}{\textbf}
\newcommand{\ra}{\rightarrow}

\begin{document}
\input{epsf}

\begin{titlepage}
\begin{center}
{\Large Search for Global Metric Anisotropy in Type Ia Supernova
Data }

\bigskip
Pankaj Jain$^a$, Moninder S. Modgil$^a$ and John P. Ralston$^b$

\bigskip
$^a${\it Physics Department, IIT, Kanpur - 208016, India}\\
$^b${\it Department of Physics \& Astronomy,}\\
{\it University of Kansas, Lawrence, KS - 66045, USA}\\
\end{center}
\vspace{1cm}

\begin{abstract}
 
We examine the Type 1a supernova data in order to determine if it 
shows any signal of large scale anisotropy.
The anisotropy is modelled by an extended G\"{o}del metric, 
which incorporates expansion along with rotation. The model is smoothly 
connected to the usual FRW type, while expressing anisotropic metric effects depending on certain parameters. We find no significant signal of anisotropy
in the data. We obtain bounds on an anisotropic redshift versus magnitude
relationship, and accompanying parameters of the G\"{o}del-Obukhov metric.
\end{abstract}

\end{titlepage}

\section{Introduction}

The large redshift type Ia Supernova data
\cite{Schmidt98,Garnavich,Perlmutter1,Perlmutter2,Riess98,Tonry,Knop,Barris}
provides direct evidence for dark energy and has played a fundamental 
role in determination of cosmological parameters. In this paper we 
examine whether the data is consistent with the assumption that the universe
is isotropic on large distance scales. There exist many observations which 
indicate a possible large scale anisotropy. These include the large scale
anisotropy in the radio polarizations from radio galaxies
\cite{Birch 1982,Jain1999}, large scale alignment of quasar optical 
polarizations \cite{huts}, alignment of CMBR quadrupole and octopole
\cite{CMB,Ralston 2004}. Here we use an anisotropic G\"{o}del type model to 
test for large scale anistropy. There is an interesting interplay between the redshift dependence and the anisotropy of the model which can be tested rather cleanly. We also test for anisotropy by using some simple extensions of the 
formula for distance modulus obtained in the standard Big Bang model. These
extensions model the effect of a wide range of anisotropic metrics in the limit
of weak anisotropy. In no
case do we find a significant signal of anisotropy. Hence we use the supernova
data to impose limits on these anisotropic models.

 The G\"{o}del universe
\cite{Godel_1949} has received significant
attention in the past few years, following the discovery
that it was an exact solution of a string theory 
\cite{Barrow 1998,Kanti 1999,Carrion 1999}. 
Indeed since its discovery in 1949, the
G\"{o}del universe attracted interest due to its fascinating
properties, including explicit rotation, the existence of closed time-like
curves (CTCs), and an anti-Machian nature.  It is well-known that G\"{o}del's
original universe is an inappropriate model because of absence of expansion
and redshift.  A number of G\"{o}del type metrics have been proposed which
allow for expansion of the universe, and in which the quandary of CTC's
are banished. In Ref. \cite{Obukhov_1988,Obukhov_2000} Obukhov generalizes G\"{o}del's model
to include expansion, listing conditions for CTC's to exist.  The revised
model contains parameters which smoothly interpolate between the usual Friedman-Robertson-Walker (FRW)model and G\"{o}del's. The independent nature of vorticity and rotation, as
discussed by Obukhov \cite{Obukhov_1988,Obukhov_2000} does not allow limits
on rotation to be placed by limits on vorticity.   Among
 distinctive predictions of the model are effects on the propagation of light.

Comparisons with observational data have been scant.  In 1982 Birch 
\cite{Birch 1982} claimed the existence of a large scale anisotropy in radio wave polarizations from extragalactic sources. He suggested that this might be observational evidence for universal rotation and anisotropy.  Birch's controversial work was revisited by many authors with conflicting outcomes 
\cite{Nodland}.  Studies dismissing the effect invariably use a statistic of {\it even} parity, while the effect itself is a corkscrew twisting or polarization with {\it odd} parity \cite{Ralston 2004}. In Ref. \cite{Jain1999},
Jain and Ralston explored tests not requiring redshift information and found significant signals of anisotropy in a large sample of data. Several other observables of radiation propagating on cosmological scales have been found to indicate a preferred direction, all of which are aligned along the same axis 
\cite{Ralston 2004,CMB,huts}. In these studies no attempt was made to identify the physical origin of anisotropy.  They may represent effects independent of gravitation and restricted to modifications of the electromagnetic sector, in which polarization observables are exquisitely sensitive.

``Dark energy'' is the popular motivation to consider models beyond the FRW metric.  Current CMB data fits a flat cosmology with $\Omega_{M}\sim  0.27$, $\Omega_{\Lambda}\sim 0.73$ for the dark matter and dark energy density, respectively.  These fits assume an isotropic universe.  Meanwhile the alignment of low angular momentum CMB multipoles, which naturally have directional features, actually {\it contradict} isotropy at a level greater than 2 sigma \cite{CMB,Ralston 2004}.  This suggests we examine the data for Type 1a supernovas for consistency with isotropy.  There is no model-independent way to go from effects on the CMB, and supernova effects in the current era, so that the two studies are independent. 
The specific  G\"{o}del-Obukhov metric that we study does not predict any
anisotropic effect on the CMB temperature due to the absence of shear in this
model \cite{Obukhov_1988}. 

It would be hard to take current supernova magnitude data and establish an effect of anisotropy without using a cosmological model.  This is because the data comes from more than one group, and is necessarily affected by uneven sampling on the sky and in redshift.  It is slightly different to ask whether the data supports isotropy.  
  An invariant correlation (defined below) shows that supernova redshifts are strongly correlated with their angular positions.  This is reasonably explained by selection.  It turns out that supernova magnitudes are even more correlated with angular position than the correlation in redshift would suggest.  
Given the difficulty of establishing an anisotropic effect, we don't even attempt to assess the importance of this signal.  The problem is that the relative importance of angular correlations of magnitudes is coupled to the model of redshift dependence, and vice-versa.  We have not seen this discussed before: it might be that by incorporating anisotropic effects, the values of  $\Omega_{M}$ or $\Omega_{\Lambda}$ would change.  Or, one might be led to place bogus limits on anisotropy hinging upon particular assumptions about the redshift dependence.  The resolution cannot be made on the basis of a single test statistic and hinges on more detailed study we will describe.  The outcome also depends on what are used for host galaxy extinctions in the supernova data, as we discuss.  

In the end our most reasonable fits do not show any signal requiring anisotropy.  Yet somewhat surprisingly, the limits on anisotropy violation parameters are comparatively weak. That is, some small anisotropy cannot be ruled out.

\section{The G\"{o}del-Obukhov Model}

The G\"{o}del-Obukhov  line element \cite{Obukhov_1988,Obukhov_2000}  is
\begin{equation}
ds^2 = dt^2 - 2 \sqrt{\sigma(t)} R(t) e^{m x_1} dt dx_2 - R^2(t) (dx_1^2
+ k e^{2mx_1}dx_2^2+ dx_3^2).
\label{eq:metric}
\end{equation}
 Let us comment on various terms: 
\begin{itemize} 
 
\item Spatial axis $x_{2}$ is singled out as breaking isotropy.  In work below, we designate the axis of anisotropy more generally as $\hat\lambda$.  
 
\item The expansion factor $R(t)$ is similar to what appears in the FRW metric. Since $R(t)$ is a single scale factor for all spatial coordinates, it also appears in the coefficient of the off-diagonal term $dt dx_2$.   

\item Other parameters in the metric are $\sigma, \, k$ and $m$. Causal 
structure of space-time is regulated by the parameter 
$k$. For $k>0$ there are no CTCs, and conversely $k \le 0$ has CTCs. 

\item The classic G\"{o}del metric is obtained by setting $R(t)=1,  \sigma(t)=1,\, m=1, \,  k=-1/2.
$

\item Parameter $\sigma$ determines the magnitude of acceleration of the 
fluid element due to rotation of the universe
 determined by the metric\cite{Obukhov_1988,Obukhov_2000}. Similarly, parameter $m$ can be identified as proportional to vorticity.  The relation of matter to the metric of course depends on the model, which has been examined in the context of Einstein's general relativity\cite{Kretchet_1988,Korotky_1993,Panov_1985} as well as the Einstein-Cartan theory\cite{Obukhov_2000}.  It is convenient to lump together the combination important for our study: the anisotropy parameter
$\tilde\rho$ defined by \ba  \tilde\rho  = \sqrt{\frac{\sigma}{k+\sigma}}. \ea  

\end{itemize}

\subsection{Magnitude Versus Redshift}

The metric Eq. \ref{eq:metric} predicts an anisotropic relationship for
magnitude as a function of redshift for standard candles.
Using the Kristian-Sachs \cite{Kristian_1966} expansion, Obukhov
\cite{Obukhov_1988} derives the following expression for the bolometric magnitude $m_{bol}$
\begin{equation}
    m_{bol}=\tilde{m}_{bol}-5  \log_{10} \left( 1+\tilde\rho
         \cos \alpha\right)\ .
	 \label{eq:mbol}
	 \end{equation}
Here $\tilde{m}_{bol}$ is the usual FRW bolometric magnitude and $\alpha$ is the angle between the source and the $x_2$ axis. If we expand the  
second term on the right hand side to leading order in $\tilde\rho$ it
gives the ``dipole'' correction due to rotation that we will bound.  
In arriving at Eq. \ref{eq:mbol} we have kept only the leading order term
in the redshift $z$ which depends
on the angular positions. This term is in fact independent of $z$. The term
linear in $z$ is found to be
proportional to $\omega_0/H_0$, which is negligibly small;  current limits 
imply $ {\omega_0\over H_0} \le 10^{-3}\ $ \cite{Obukhov_1988,Birch 1982}.
Similarly, we ignore tiny kinematic effects from our reference frame's velocity relative to the rest frame of the CMB. If
$\tilde\rho<<1$, which indeed must be the case,
we may expand the logarithm and keep only the leading order term in
$\tilde\rho$.

\subsection{Relation to Distance Moduli}

The distance modulus $\mu_p$ of a supernova
is defined by $\mu_p = m - M$, where $m$ and $M$ are the apparent and
absolute magnitudes respectively. This is related to the luminosity
distance $d_L$ by the relationship
\begin{equation} \mu_p = 5\log_{10} d_L +25
\label{eq:dL}\end{equation}
We parametrize the luminosity distance in the FRW limit, setting
$\sigma=0$, as follows \cite{Visser}
\begin{equation}
    d_L={cz\over H_0}\left[1+\frac{1}{2}(1-q_0) z
    -\frac{1}{6} (1-q_0-3q_0^2+j_0) z^2 \right]
\label{eq:dL_FRW}
\end{equation}
where $q_0$ and $j_0$ are respectively the deceleration and
the jerk parameters, defined in Ref. \cite{Visser}. We can now express the
distance modulus in the G\"{o}del-Obukhov metric, Eq. \ref{eq:metric},
as follows:
\begin{eqnarray}
    \mu_p&=&\mu_{p0}+5 \log_{10} z + 5 \log_{10} \left[1+\frac{1}{2}(1-q_0) z
    -\frac{1}{6} (1-q_0-3q_0^2+j_0) z^2 \right]\nonumber\\
     &-& 5 \log_{10} \left[1+ \tilde\rho  \cos \alpha\right], 
\label{eq:mu_p_Godel}
\end{eqnarray}
where
\begin{equation}
    \cos \alpha = \hat{\lambda} \cdot \hat{r} . 
\end{equation}
Here $\hat{r}$ is the direction of the source, and (to repeat) $\hat{\lambda}$ is the preferred axis of anisotropy of the metric.  We have included an overall parameter $\mu_{p0}$ to allow for a zero point correction, which includes uncertainty in the Hubble
parameter.

\begin{figure}
\epsfbox{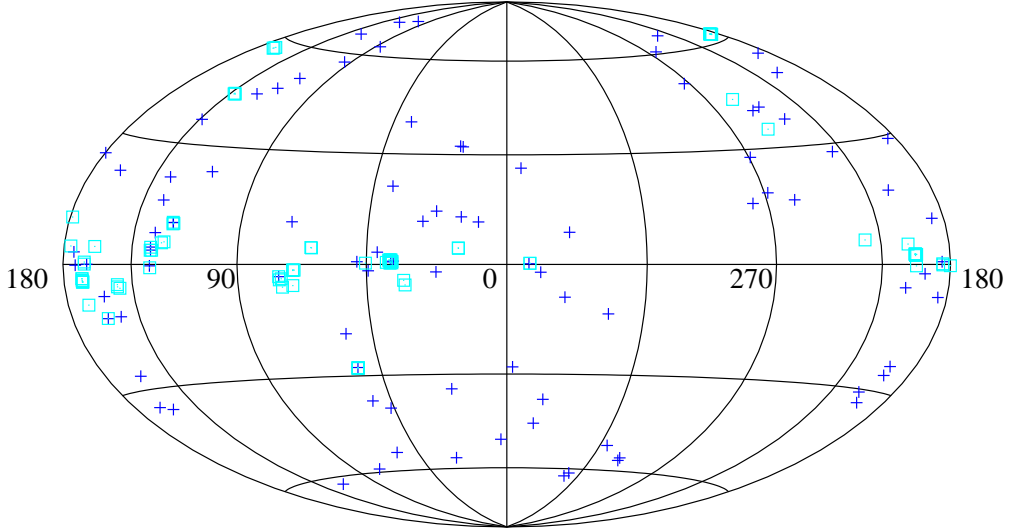}
\caption{Scatter plot of angular positions of the 186 object
gold and silver set compiled in Ref. \cite{Riess_2004} in Equatorial
J2000 coordinate system.  Symbols indicate low redshifts
($z\le 0.4$, plusses) and high redshifts ($z>0.4$, squares). }
\label{fig:scatter}
\end{figure}

\section{Angular Correlations}

Here we test for signals of angular dependence
in the type 1a supernova data. We examine
the ``gold'' data set comprising 157 sources from Ref. \cite{Riess_2004}, 
as well as the gold and ``silver'' data set of 186 sources.   A scatter plot of the data is shown in
Fig. \ref{fig:scatter}.   One sees that the data population is quite unevenly distributed in the sky,  particularly at high redshifts.  There is nothing remarkable about this distribution, of course.  

In many statistical tests the sample selection might cause a spurious signal of angular dependence.  Hence naive signals of anisotropy must be interpreted with care.  Conversely, selection effects may also mask a true signal of angular dependence that might be present. We now turn to tests of correlation that do not depend on the sample population. 

\subsection{JM Correlation Statistics}

We quantify correlations of distance moduli with angular
positions on the sky by using the Jupp-Mardia \cite{Jupp} (JM) angular correlation coefficient $r$.  In general, one maps angular variables into ``vectors'' (upper case
$X_{i} $) that transform linearly under a change of coordinates. From the map statistics which are rotationally invariant are constructed.  

The procedure uses correlation matrices 
\ba C_{XX;  \,ij}= <(X-\bar X)_{i}(X-\bar X)_{j}>;  \:\:
C_{XY; \, ij}= <(X-\bar X)_{i}(Y-\bar Y)_{j}> , \nn \ea etc. 
where $< \,>$ indicates a summation over the data set and $\bar X$, $\bar Y$
refer to the mean values over the sample.  
Explicitly $<X_{i}X_{j}>=\sum_{\alpha} \, X^{\alpha}_{i}X^{\alpha}_{j}, $ where $X^{\alpha}_{j}$ is the $jth$ vector component of the $\alpha$-th data point.   The general JM statistic $r^{2}(X,\, Y)$ is the  combination 
\begin{equation}
r^{2}(X,\, Y) = tr(\,  C_{XX}^{ -1}C_{XY} C_{YY}^{-1} C_{YX} \,) , \label{JM} 
\end{equation} 
where matrix multiplication is implied, $C^{-1}$ denotes an inverse matrix, and $tr$ indicates the trace. 

The JM statistics are a very general test of correlation of the underlying distribution $f(X, \, Y)$.  Distributions are {\it uncorrelated} if $$ f(X, \, Y)=f(X)f(Y), \:\:\:\: (null \, model), $$ where $f(X)$ is the marginal distribution of $X$, and similarly for $Y$.  For a sample of size $n$, JM show that 
the product $nr^{2}$, where $r^2$ is the statistic defined in Eq. \ref{JM},
from an uncorrelated null distribution is distributed like $\chi_{\nu}^{2}$, where $\nu$ is the total number of components of $X$ and $Y$.  Extensive Monte Carlo calculations confirmed this distribution in 
Ref. \cite{Jain1999}.

It is very important that JM test statistic is independent of the particular marginal distributions. Population (marginal distribution) effects are automatically removed by inverse matrices used in the definition. For our purposes we assign $\hat r \ra X$, where $\hat r$ is the unit vector of the position on the sky in equatorial coordinates.  We assign scalar quantities such as distance moduli $\mu_p \ra Y$.  Eq. \ref{JM} applies without change, the meaning of $C(YY)$ simply becoming the sum of squares of $Y$, and so on.   By construction (and previous consistency tests) the uneven sampling of supernovas on the sky cannot create a significant false correlation of $nr^{2}$ if the sample distribution obeys the null model.  Thus the JM test statistics do not require any particular model of anisotropy, but they can rule out a null hypothesis of isotropy.  The following facts are found:

\begin{table}
  \centering

  \begin{tabular}{cccc}
\hline
 $ sample$  & $n$  &  $n r^{2} $ & $P-value$  \\
 gold \,($z$)  & 157  & 20.4 & $1.4 \times 10^{-4}$   \\
  gold \,($\mu$)  & 157  & 27.1 & $5.6 \times 10^{-6}$   \\
    gold \,($A_{V}$)  & 133  & 15.5 & $ 1.4\times 10^{-3}  $   \\
 gold and silver \, ($z$)   & 186  & 16.3 &  $9.8 \times 10^{-4}$ \\
  gold and silver \, ($\mu$)   & 186   & 23.5 &  $3.2\times 10^{-5}$ \\
    gold and silver \, ($A_{V}$)   & 156   & 9.5 &  $2.4\times 10^{-2}$ \\
\hline
\end{tabular}
  \caption{ Jupp-Mardia invariant statistics $r^{2}$ for the correlation of redshifts ($z$),  distance moduli ($\mu$), and host extinctions ($A_{V}$) of supernovas with their angular position on the sky.  Symbol $n$ is the sample size, and the $P$-value is the probability to see the statistic or larger ones in an uncorrelated null distribution. Supernovas for which host extinction values are not
  available are removed while computing the statistics for $A_V$. }
\label{tab:JMcor}
\end{table}

\begin{itemize} 
 
\item {\it Supernova Magnitudes Break Isotropy: }  Evaluation of the JM correlations produced Table \ref{tab:JMcor} comparing $n r^{2}(z)$ and $n r^{2}(\mu)$.  As discussed in the introduction we cannot expect supernova sampling to be uniform in ``depth'', namely redshift.  This is reflected in the large  
correlation of redshift with angular position, which indicates sampling 
effects. Meanwhile in both the gold set, and the gold and silver set, {\it the magnitudes themselves are much more correlated than the redshifts.}  This cannot be attributed to sample skewing by the Milky Way: the numbers represent a {\it correlation} of magnitudes with position, not the distribution of the positions themselves.  Nor can the results be explained away by extinction effects of our galaxy, unless those effects have been estimated incorrectly.   Our own galaxy's extinctions have already been removed by the observers.  We are not sure how to assess the relative probability of the magnitude correlation given the redshift correlation. For reference, the ratio of $P-values$ is of order $0.03-0.04$, perhaps a significant (``2-$\sigma$'') effect. 

\item  {\it Host Galaxy Extinctions Break Isotropy: } In Ref. \cite{JR05} evidence was found for evolution or bias in the host galaxy extinctions of supernova in the gold set.  The trend is that host extinctions are strongly correlated with their deviation from the Hubble plot.  The method by which extinctions are assigned by observers tends to produce the same signal as acceleration.  We decided to compute the JM correlation of host extinctions with angular positions (Table \ref{tab:JMcor}). For the gold 
set the result is\footnote{ 24 sources for which extinctions are not reported have been deleted.} \ba nr^2(A_{V}) = 15.5, \nn \ea a $P$-value of $1.4\times 10^{-3} $.  This ``$3 \sigma$''-effect is statistically quite significant.  Perhaps the combination of effects explains the pattern seen in Table \ref{tab:JMcor}.  Admittedly there is no reason {\it a priori} for the two effects to be oriented in a way to add.  Nor can this possibility be checked with the $JM$ correlations themselves. 

\item  {\it Anisotropy of Magnitudes and Host Galaxy Extinctions Are Related: }  Comparing the correlation of host extinctions and magnitudes requires a directional statistic.  We computed $Y$-weighted average unit vectors $<Y \hat X >$, \ba  <Y  \hat X > = (\, \sum_{\alpha} \, (Y^{\alpha} -\bar Y) \vec X^{\alpha} \,)_{normalized}.  \nn \ea  Again the sum is over galaxy labels $ \alpha$. Subscript -$normalized$ indicates that result is normalized after making the sum.  
This statistic has two degrees of freedom, and like the $JM$ correlations its direction {\it is not} strongly affected by the marginal distribution on the dome of the sky. By construction $<Y \hat X >$ vanishes if distributions are uncorrelated, and any bias in the $\vec X$ marginal distribution should average symmetrically to zero in an uncorrelated null distribution. Moreover, the relationship of $<Y \hat X >$ calculated for $Y=z$ and for $Y=A_{V}$ should be random if the host extinctions are not causing the extra angular correlation of magnitudes seen in Table \ref{tab:JMcor}.  The naive probability of overlapping one unit vector with another isotropically is 
$P( <Y_1 \hat X >, \, <Y_2 \hat X >)  = (1- <Y_1 \hat X > \cdot  <Y_2 \hat X >)/2.$  We only conduct this study for the gold set and those entries for which $A_{V}$ are listed. Surprisingly all the $Y$-weighted vectors line up, with:

\begin{itemize} 

\subitem $P(  < z \hat X >, \, < \mu  \hat X >)=  2.5  \times 10^{-2} $  

 \subitem $P(  <A_{V} \hat X >, \, < z \hat X >)= 6.9 \times 10^{-2}$
 
  \subitem $P(  <A_{V} \hat X >, \, < \mu  \hat X >)= 1.2  \times 10^{-2} $

\end{itemize} 

One of these statistics is dependent; as already discussed the $z- \mu$ coincidence is also perfectly expected.  The direction of $<A_{V} \hat X >$ is 
$RA=114^{o},$ $DEC =8^{o}$.  It is hard to understand how the host extinctions might have become correlated\footnote{Using the rescaled extinctions of Ref. \cite{JR05} will not change the result} in this way.
However, given the fact that this angular correlation exists, the higher correlation of magnitudes reported relative to the correlation in $z$ seems to be explained.  
\end{itemize}

{\it Up-Down Test:}  Since the JM tests are somewhat elaborate we also used a simple null test for isotropy. Fig. \ref{fig:scatter} shows that the data naturally split into two separate groups: a group lying in the region $100\le RA \le 250$ (85 sources, set 1), and another group (72 sources, set 2). For the test we fit the isotropic cosmological
parameter $\Omega_M$ separately for each set, constraining $\Omega=\Omega_M + \Omega_V=1$.  The best fit for set 1 gives $\Omega_M^{(1)} = 0.29\pm 0.05$, while set 2 gives $\Omega_M^{(2)} = 0.33\pm 0.05$.  This test yields no evidence for a systematic difference between the two sets. 

However it is reasonable to make a search over different directions defining hemispheres labeled ``up'' and ``down'', fitting $\Omega_{M-up}$ and $\Omega_{M-down}$, in obvious notation.  Such a search introduces two additional parameters, the orientation of a $z$-axis defining ``up''. This test finds a maximum difference $\Omega_{M-up}-\Omega_{M-down} = 0.14, $ with 
the axis position ${\rm RA} = 307^o$, ${\rm Dec} = -20^o$, roughly orthogonal
to both celestial and galactic poles. 
Given that the one standard deviation error $\Delta \Omega_M= 0.05$, and taking into account the 2 axis parameters, this
difference is statistically significant at the 5 \% level. That is, 
there is only 5 \% probability that such a difference may arise by a 
fluctuation in an isotropic sample.  

While blunted by the $\Omega_{M}$ fitting procedures, the two-hemisphere test indicates a weak signal of anisotropy.  This is quite consistent with its crudeness and lower sensitivity compared to the JM test. 

\paragraph{Model-Independent Correlation Summary} 

To summarize results up to here: The supernova data are hardly consistent with isotropy when examined with the model-independent angular correlation tests.  At the same time there is at most a modest signal of fundamental anisotropy, 
which might be brought out further if the G\"{o}del-Obukhov model is 
appropriate.

\subsection{ $\chi^{2}$ Fits}

Here we fit data to the G\"{o}del-Obukhov anisotropic cosmological model,
Eqs. \ref{eq:metric}-\ref{eq:mu_p_Godel}.

\begin{table}
\centering
\begin{tabular}{lccccc}
\hline
$data \, set$  &    $ q_{0}$ & $j_{0}$ & $\tilde\rho$
&$ \chi^{2}$ & $P$  \\
\hline
gold and silver - best fit      & $-0.77$ & 1.96 &    0.02  &   228.4 & 0.008 \\gold and silver     &$-0.70$  & 1.60 &    $\equiv 0$  &   230.1 & 0.010  \\
gold - best fit &  $-0.68$ &1.50   & 0.021 & 174.5  & 0.092     \\
gold  &   $-0.63$ & 1.23  & $\equiv 0$  & 176.1  &  0.11    \\
gold and silver - $\delta$-best fit      & $-0.53$ & 0.72 & 0.029  & 205.5
& 0.085   \\
gold and silver - $\delta$     & $-0.51$  &0.57  & $\equiv 0 $ & 209.2 & 0.082\\gold - $\delta$ -best fit  & $-0.43 $ & 0.28  & 0.031 & 152.4 &0.43 \\
gold - $\delta $  & $-0.45 $  & 0.29 & $ \equiv 0$ & 155.3  & 0.43     \\
\hline
\end{tabular}
\caption{The $\chi^2$ values for the gold and gold-silver set. Results are
given for both the fit to the FRW model ($\tilde\rho\equiv 0$) and the
anisotropic Godel-Obukhov model. We also give results using the corrected
extinction values, with the rescaling parameter $\delta=-0.43$. }
\label{tab:chi2}
\end{table}

\subsubsection{Nominal Extinctions} 
In this study we use the extinction values published by Riess et al\cite{Riess98}.  
Parameters are fit by minimizing $\chi^{2}$, defined by
\begin{equation}
\chi^2 = \sum_i  \,  {(\, \mu_p^i - \mu_0^i \,)^{2} \over(\delta\mu_0^i)^{2}},
\end{equation}
where $\mu_0^i$ is the observed distance moduli, $\mu_p^i$ is the theoretical
prediction (Eq. \ref{eq:mu_p_Godel}) and $\delta\mu^i_0$ are the reported 
errors. 
We vary six parameters:  three FRW parameters $\mu_{p0}$, \, deceleration ($q_0$) and jerk ($j_0$), the anisotropy parameter $\tilde\rho$, and the two axis
parameters. Results are summarized in Table \ref{tab:chi2}.  The minimum $\chi^2$ in the rotating model
for the gold and silver set is found to be 228.4 for $\tilde\rho=0.02$.  Given
six parameters this gives a $\chi^2/dof =1.27$, where $dof$ stands for
the degrees of freedom.  This is to be compared with the $\chi^2= 230.1$ in the FRW model, which has three parameters, and $\chi^2/dof =1.26$.  The rotating model provides a slightly worse overall fit to the data.  Yet both fits are poor: the probability is quite low for a good fit to give such $\chi^{2}/dof$ values for so many data points. $P$-values listed in Table \ref
{tab:chi2} are the probability to see the
statistic or larger in $\chi_{\nu}^{2}$, where $\nu$ is the number of degrees of freedom.

\begin{figure}
\epsfbox{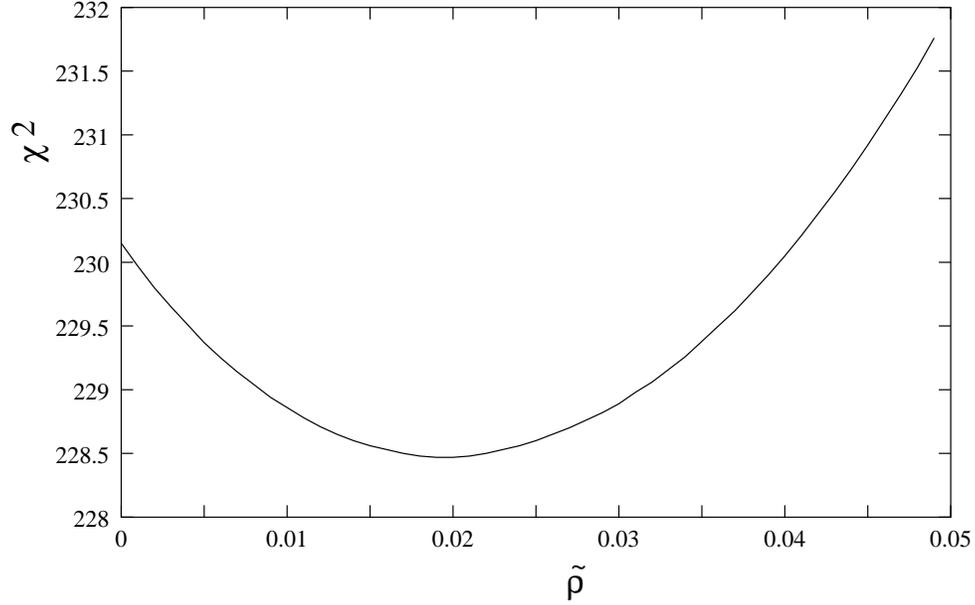}
\caption{$\chi^2$ for the 186 object data set for the
 G\"{o}del-Obukhov metric as a function of the parameter $\tilde\rho$.
}
\label{fig:chi2}
\end{figure}

Dependence of $\chi^2$ with $\tilde\rho$ is shown in Fig. \ref{fig:chi2}.  From the fit we obtain a limit on the parameter $\tilde\rho$: one
standard deviation limits are 
\begin{eqnarray}
 \tilde\rho &<& 0.046 \, ({\rm gold\   and\  silver}) ; \nonumber \\
 \tilde\rho &<& 0.049  \, ({\rm gold})\ .
\end{eqnarray}

\subsubsection{Corrected Extinctions}
 
Jain and Ralston \cite{JR05} found that re-scaling the host extinction values
generates a much better fit to the conventional isotropic model, decreasing the
$\chi^{2}$ values by more than 22 units, and bringing $\chi^{2}/dof$ close to unity.  Re-scaling also eliminates an alarming
correlation between residuals and extinctions.  The re-scaling correction
replaces extinction values $A_{V}$ by the rule
\ba A_V \rightarrow A_V(\delta)  = (1+\delta) A_V \label{ext}, \ea with $\delta = -0.43$.
Using the corrected $A_V(\delta) $ values, the $\chi^2$ values for the gold set
(Table \ref{tab:chi2}) have $\chi^2/dof =1.01$. These $\chi^2$ values are
slightly different from that
obtained in Ref. \cite{JR05}, since here we include
the deceleration and jerk $j_0$ parameters, instead of
$\Omega_M$ and $\Omega_\Lambda$ and the constraint
$\Omega_M + \Omega_V = 1$ is not imposed. Values for the gold and silver set 
with various assumptions are given in Table \ref{tab:chi2}.

With corrected extinctions the fit to angular dependence is just as good as the fit to the FRW metric.  For the Gold data set the axis position is 
${\rm RA} = 229.5^o$ and 
${\rm Dec} = -53.2^o$, roughly orthogonal to the galactic pole. 
The axis position for
the Gold and Silver set is ${\rm RA} = 199.6^o$ and
${\rm Dec} = -47.5^o$.
We determine the upper limit of the parameter
$\tilde\rho$ by demanding that $\chi^2$ for the anisotropic model is at most
one unit larger than that in the FRW model, yielding
\begin{eqnarray}
 \tilde\rho &<& 0.066 \, ({\rm gold}  \ {\rm and}  \ {\rm silver}) ; \nonumber \\
 \tilde\rho &<& 0.067  \, ({\rm gold})\ .
\end{eqnarray}

\subsubsection{Quadrupole Model}

Up to here we have considered bounds on a specific model which incorporates
rotation along with expansion. Yet the bound is really applicable to any
model which at leading order gives a dipole contribution, $\hat \lambda\cdot
\hat r$, where $\hat\lambda$ is a preferred axis.

A natural generalization of the model would incorporate a quadrupole contribution.  We restrict our study to dependence going like $\cos^2\alpha$ , while allowing the parameters of the
preferred axis to vary.  The anisotropic correction to the distance
modulus is then given by
\begin{equation}
\delta \mu_p = -5  {\rho' \cos^2 \alpha \over \ln(10)}
\end{equation}
where $\rho'$ can be positive or negative. In order to compare this
with the distance modulus in the G\"{o}del-Obukhov model we may expand the 
last term in Eq. \ref{eq:mu_p_Godel} to leading order in $\tilde \rho$. 
For the gold and silver set
the minimum $\chi^2=225.1$, $\chi^2/dof=1.25 $ with parameters
$q_0=-0.68$, $j_0=1.5$ and $\rho'=0.078$.
The fit obtained here is slightly better than that obtained in the FRW model.
The previous bound becomes considerably looser, yielding $-0.10<\rho'<0.17$.
If we use the extinction correction we find $ -0.14 < \rho'< 0.11  $. The minimum
$\chi^2=206.1$, $\chi^2/dof =1.145 $ with parameters
$q_0=-0.585$, $j_0=0.949$ and $\rho'= - 0.060$.

\subsection{Possible Selection Effects}

As seen in fig. 1 the data  does not have
 a uniform angular distribution.  Nor does one expect such a distribution given
 the selection effects of observations.  The redshift values also show a
  correlation with angular position due to selection effects.  We next ask
the question how these effects might change our limits.

  We find that the  selection effects do not affect our limits.
  Although selection effects might certainly generate an artificial signal of anisotropy, at least for some observables, the converse of hiding a true signal and showing a false null is a much different question. Much depends on how the analysis is arranged.  In the present case these can generate neither a spurious signal nor a spurious null.

   We demonstrate this with a simulation. We generated a simulated data set containing 400 sources. We arbitrarily assign an error of 0.2 to each simulated distance
    modulus. The data is generated from a gaussian distribution with the
     mean value at any redshift given by the distance modulus obtained from the
      FRW metric. For 200 sources we uniformly assign a redshift range between
       0 and 1. For the remaining 200 sources we assign redshift values uniformly
        in the range 0 to 2. These two sets are assigned two
	 different, arbitrarily chosen, angular positions. This of course is an egrregiously selected data set.

	 We fit the simulated data with our model for distance modulus seeking angular dependence. The analysis  shows no significant signal of anisotropy.   Despite the
	  fact we have taken a highly anisotropic sample distribution on the sky, and
	   a set that shows a very large angular redshift dependence, the {\it correlation} of distance modulus and angular position is unaffected.

	   We carried out several examples of such simulations with different choice of angular positions and with data scattered at more than two positions on the sky. In no case
	    do we find any significant signal of anisotropy.

	    This shows that our tests of the correlation of the underlying distributions do not respond to substantial bias in the {\it marginal} distributions.   In order for selection on the {\it marginal} distributions to conceal an actual correlation, one would have to deliberately cut out (select away) those special data capable of showing correlation.  The only way to do so without cheating would be to select a very tiny range of redshifts so that the lever arm to find correlation would be too small. Such a cut will only make our limits more 
conservative. However they will still be rigourously applicable.

\section{Conclusions}

We have examined the magnitudes of large redshift type 1a supernova data for consistency with the G\"{o}del-Obukhov model. Using the nominal extinction coefficients gives poor fits, in which anisotropy makes the fit worse.  Using corrected extinction coefficients gives excellent fits, while also opening up the allowed region of the anisotropy parameter.  
We have also tested for dipole and quadrupole anisotropy in the supernova
data. In none of the cases have we found strong evidence in favor of any anisotropic effects. It is nevertheless impossible to rule out anisotropy entirely. 

{\it Acknowledgments: Research supported in part under DOE Grant Number
DE-FG02-04ER14308.}

\end{document}